\documentstyle[11pt,aaspp4]{article}

\slugcomment{Submitted to Astrophys. J. Lett.}

\begin{document}

\title{Chemical Abundances and the Metagalactic Radiation Field at High 
Redshift}

\author{Daniel Wolf Savin}
\affil{Columbia Astrophysics Laboratory, Columbia University, \\ 
New York, NY 10027, USA}
\authoremail{savin@astro.columbia.edu}

\begin{abstract} 

We have carried out a series of model calculations of the photoionized
intergalactic medium (IGM) to determine the effects on the predicted
ionic column densities due to uncertainties in the published
dielectronic recombination (DR) rate coefficients.  Based on our
previous experimental work and a comparison of published theoretical DR
rates, we estimate there is in general a factor of 2
uncertainty in existing DR rates used for modeling the IGM.  We
demonstrate that this uncertainty results in factors of $\sim 1.9$
uncertainty in the predicted N~V and Si~IV column densities, $\sim
1.6$ for O~VI, and $\sim 1.7$ for C~IV.  We show that these systematic
uncertainties translate into a systematic uncertainty of up to a factor
of $\sim 3.1$ in the Si/C abundance ratio inferred from observations.
The inferred IGM abundance ratio could thus be less than (Si/C)$_\odot$
or greater than 3(Si/C)$_\odot$.  If the latter is true, then it
suggests the metagalactic radiation field is not due purely to active
galactic nuclei, but includes a significant stellar component.  Lastly,
column density ratios of Si~IV to C~IV versus C~II to C~IV are often
used to constrain the decrement in the metagalactic radiation field at
the He~II absorption edge.  We show that the variation in the predicted
Si~IV to C~IV ratio due to a factor of 2 uncertainty in the DR rates is
almost as large as that due to a factor of 10 change in the decrement.
Laboratory measurements of the relevant DR resonance strengths and
energies are the only unambiguous method to remove the effects of these
atomic physics uncertainties from models of the IGM.

\end{abstract}


\section{Introduction}
\label{sec:Introdution}

Many fundamental question of cosmology can be addressed through
observations of the Ly-$\alpha$ forest.  For example, observation of
metal absorption lines can be used to constrain the spectral shape and
history of the metagalactic radiation field, the chemical evolution of
the universe, and the initial mass function (IMF) of the earliest
generation of stars (Songaila \& Cowie 1996; Giroux \& Shull 1997;
Boksenberg 1998; Songaila 1998).  Interpreting spectra from the
Ly$\alpha$ forest is carried out using both single phase models
(Giroux \& Shull 1997; Songaila 1998) and cosmological models of the
IGM employing semi-analytic approximations or hydrodynamical
simulations (Miralda-Escud\'e et al.\ 1996; Bi \& Davidsen 1997;
Hellsten et al.\ 1997; Rauch, Haehnelt, \& Steinmetz 1997; Zhang et
al.\ 1997; Gnedin \& Hui 1998; Riediger, Petitjean \& M\"uckert 1998;
Madau, Haardt, \& Rees 1999).  These various models use different
approximations and assumptions.  However, one thing they all have in
common is the need to calculate the ionization structure of the
photoionized IGM.  This is typically carried out using plasma codes
which are written specifically for modeling the ionization structure of
photoionized gas.  One of the most commonly used codes for this purpose
is CLOUDY (Ferland et al.\ 1998).

Fundamental to the accuracy of these plasma codes and any inferred
astrophysical conclusions is calculating the correct ionization
balance.  This in turn depends on the accuracy of the dielectronic
recombination (DR) rates at IGM temperatures ($\sim 10^4$ K).  At these
temperatures DR is the most important electron-ion recombination
process for almost all ions (Arnaud \& Rothenflug 1985; Arnaud \&
Raymond 1992; Kallman et al.\ 1996).

In this Letter we demonstrate that uncertainties in the DR rates for
C~IV, N~V, O~VI, and Si~IV signficantly hamper our ability to constrain
reliably the chemical abundances and the shape of the metagalactic
radiation field at high redshift.  In Sec.~\ref{sec:DR} we review the
status of the relevant DR rates and their uncertainties.  The model we
use to calculate the ionization structure of the IGM is presented in
Sec.~\ref{sec:Model}.  In Sec.~\ref{sec:Astro} we present the results
of our simulations, demonstrate the effects of the estimated
uncertainties in the DR rates, and discuss the astrophysical
implications.  We present our conclusions in
Sec.~\ref{sec:Conclusions}.

\section{Dielectronic Recombination}
\label{sec:DR}

The lack of reliable DR rates is the dominant uncertainty in ionization
balance calculations of photoionized plasmas (Ferland et al.\ 1998).  A
critical evaluation of published theoretical DR rates suggests a factor
of 2 or more uncertainty is inherent in the different theoretical
techniques used to calculate DR for ions with partially filled valence
shells (Arnaud \& Raymond 1992; Savin et al.\ 1997, 1999).  This is
supported by laboratory measurements which have turned up errors of
factors of 2 to orders of magnitude in calculated DR rates (Linkemann
et al.\ 1995; Savin et al.\ 1997, 1999).  The measurements also
demonstrate that it is not possible a priori to know which set of
calculations, if any, will agree with experiment.  Taken all together,
these results suggest that, for ions with partially filled valence
shells, a factor of 2 uncertainty exists in almost all theoretical DR
rates currently used for modeling photoionized plasmas.

\begin{figure}
\plotone{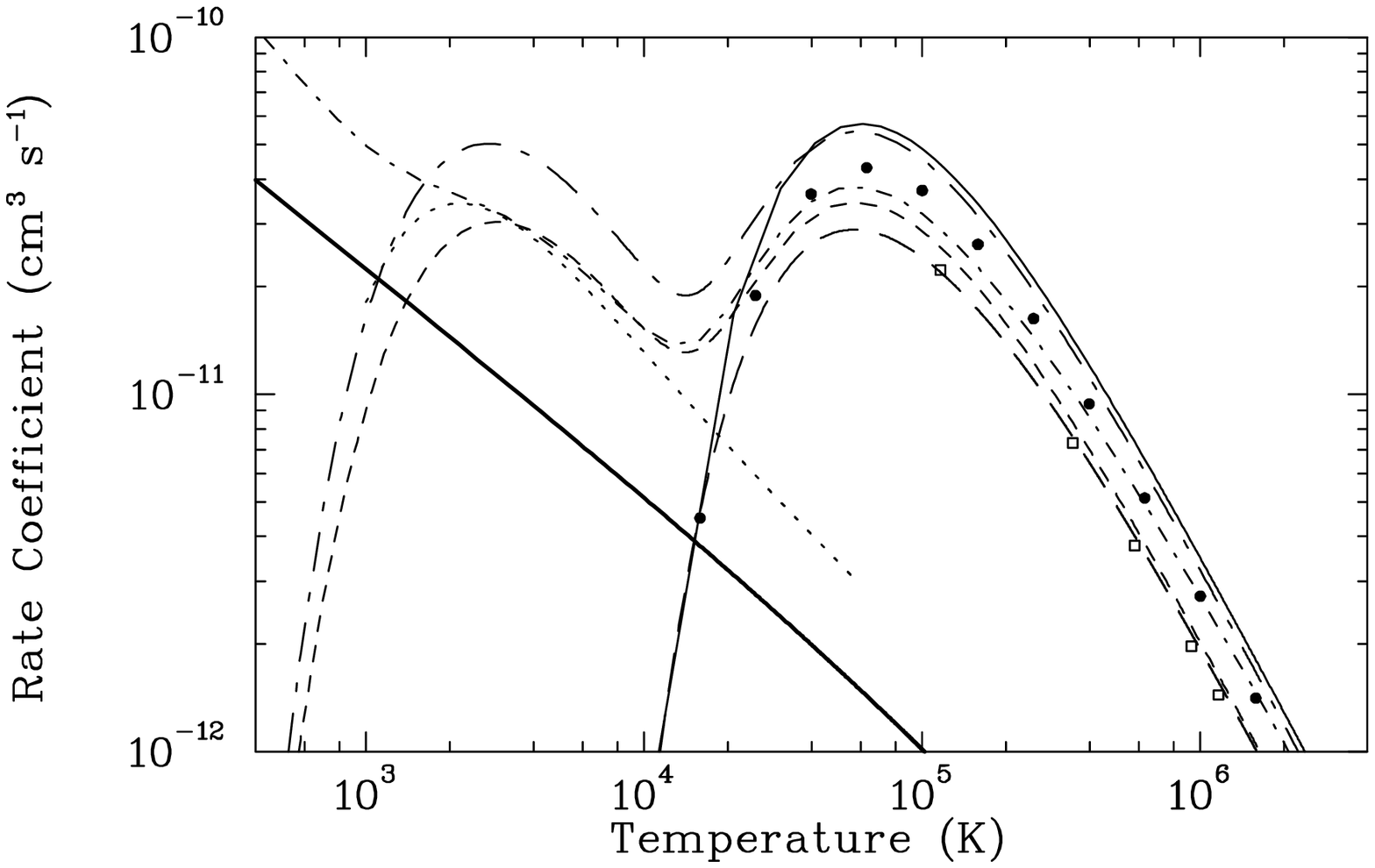}
\caption{Published theoretical C~IV to C~III DR rates versus electron
temperature.  Calculations are from Burgess (1965, thin solid line),
Shull \& van Steenberg (1982, long-dashed curve); Nussbaumer \& Storey
(1983, short-dashed curve); McLaughlin \& Hahn (1983, medium-dashed
curve); Romanik (1988, dotted-long-dashed curve); Badnell (1989, filled
circles); Chen (1991, open squares); and Nahar \& Pradhan, who
calculated a combined radiative recombination (RR) and DR rate, (1997,
dotted-medium-dashed curve).  The thick solid curve is the RR rate from
P\'equignot, Petitjean, \& Boisson (1991).}
\label{fig:CIV}
\end{figure}

As an example, we show in fig.~\ref{fig:CIV} the published theoretical
rates for DR onto C~IV.  Consisting of one electron outside of a closed
shell, C~IV is one of the simplest ions to treat theoretically and
there have been numerous DR calculations, but theory has yet to
converge.  There is still a factor of $\sim 2$ spread between the
different calculations over the entire temperature range.

Laboratory measurements are needed to determine the true DR rates and
the best theoretical techniques for calculating DR.  But as
demonstrated by Savin et al.\ (1999), it is not possible to distinguish
between different theoretical techniques based solely on the comparison
of rate coefficients with experiments.  The only unambiguous way to
benchmark DR theory is through a detailed comparison of resonance
strengths and energies.

N~V, O~VI, and Si~IV are similar to C~IV in that they consist of one
electron outside of a closed shell.  Based on our experimental studies
and theoretical comparisons, we estimate a factor of 2 uncertainty in
the calculated rates for DR onto N~V, O~VI, and Si~IV.  DR onto C~IV
has recently been measured by Mannervik et al.\ (1998) and Schippers
(1999) and his collaborators.  These groups are working to generate new
C~IV DR rates.
 
\section{Model}
\label{sec:Model}

Hellsten et al.\ (1998) have carried out hydrodynamic cosmological
simulations for a redshift of $z=3$.  They present the resulting
relationships for electron temperature $T_{\rm e}$ versus total
hydrogen density n$_{\rm H}$ and for n$_{\rm H}$ versus H~I column
density N$_{\rm HI}$.  We use their results, along with CLOUDY version
90.05, to investigate the effects of the uncertainty in the C~IV, N~V,
O~VI, and Si~IV DR rates on the predicted IGM column densities for
these ions.  The temperature-density relation depends partly on the
ionization structure of the gas and hence on the DR rates used.  To
simulate the possible effects the DR uncertanities have on this
relation, we have also carried out calculations with $T_{\rm e}$
increased and decreased by a factor of 2.  This does not significantly
affect the conclusions in this paper.

We use the same spectral shape for the metagalactic radiation field as
Hellsten et al.\ (1998) but have varied the decrement at the He~II
absorption edge (4 Ryd) by factors of 1, 2, 10, and 100.  We assume
that the decrement at the 4 Ryd does not affect the temperature-density
relationship.  This assumption is not strictly valid.  Hui \& Gnedin
(1997) have shown that a decrement of $10^4$ (twice our maximum
decrement) does decrease the temperature, but by less than a factor of
2.  Our modeling shows this uncertainty in the temperature-density
relationship does not significantly affect the conclusions in this
paper.

We use a flux at 912~\AA\ of $J_\nu = 10^{-21}$ erg cm$^{-2}$ s$^{-1}$
sr$^{-1}$ Hz$^{-1}$.  For a metallicity, we use $[{\rm Z/H}] \equiv
\log(n_Z/n_H)-\log(n_Z/n_H)_\odot = -2$.  At N$_{\rm H} \gtrsim 10^{17}$
cm$^{-2}$, the IGM begins to become optically thick and self-shielding
of the UV radiation needs to be taken into account (Rauch, Haehnelt, \&
Steinmetz 1997).  Here we restrict our calculations to N$_{\rm H}
\le 10^{17}$ cm$^{-2}$.

\section{Simulations and Astrophysical Implications}
\label{sec:Astro}

\begin{figure}
\plotone{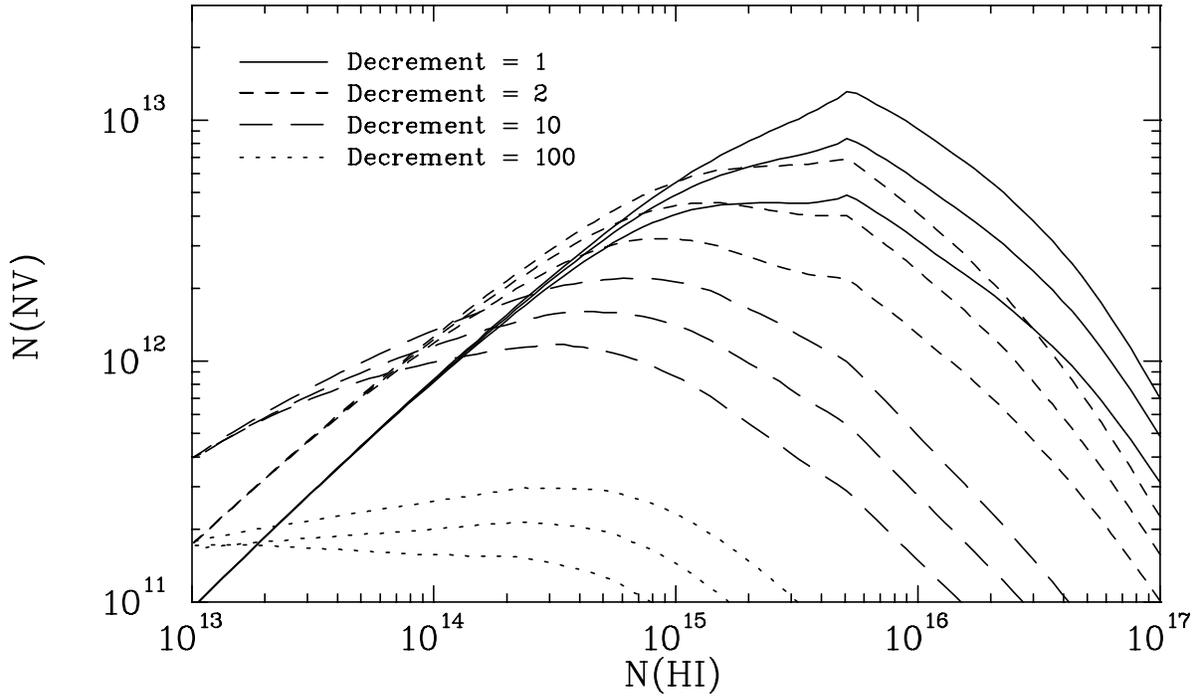}
\caption{Predicted N~V column density versus H~I column density for the
model described in Sec.~\ref{sec:Model}.  Each set of three curves
represent a metagalactic radiation field with a decrement at 4 Ryd of 1
(solid curves), 2 (short-dashed curves), 10 (long-dashed curves), and
100 (dotted curves).  We have also varied the the N~V to N~IV DR rate
and left the other rates unchanged.  For each set of three curves, the
result are shown with the rate decreased by a factor of 2 (upper
curve), unchanged (middle curve), and increased by a factor of 2 (lower
curve).}
\label{fig:NVvHI}
\end{figure}

\begin{figure}
\plotone{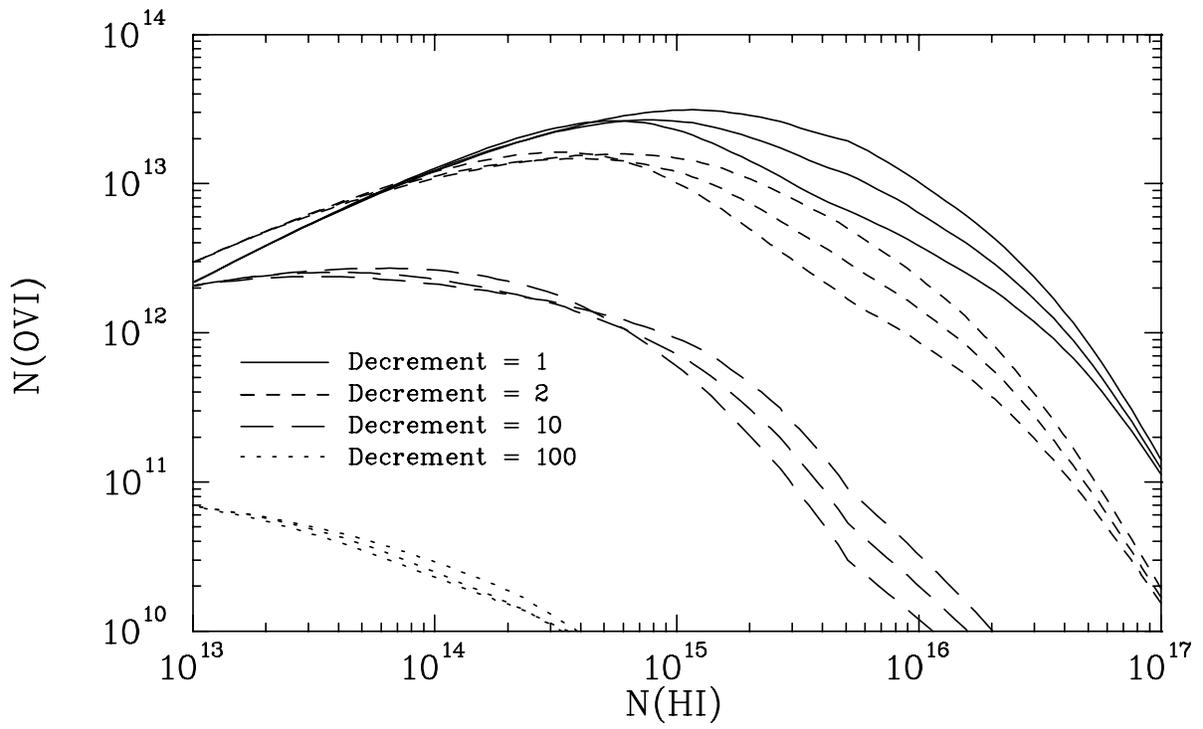}
\caption{Predicted O~VI column density versus H~I column density.  We
have varied the the O~VI to O~V DR rate and left the other rates
unchanged.  See Fig.~\ref{fig:NVvHI} for further
details.}
\label{fig:OVIvHI}
\end{figure}

\begin{figure}
\plotone{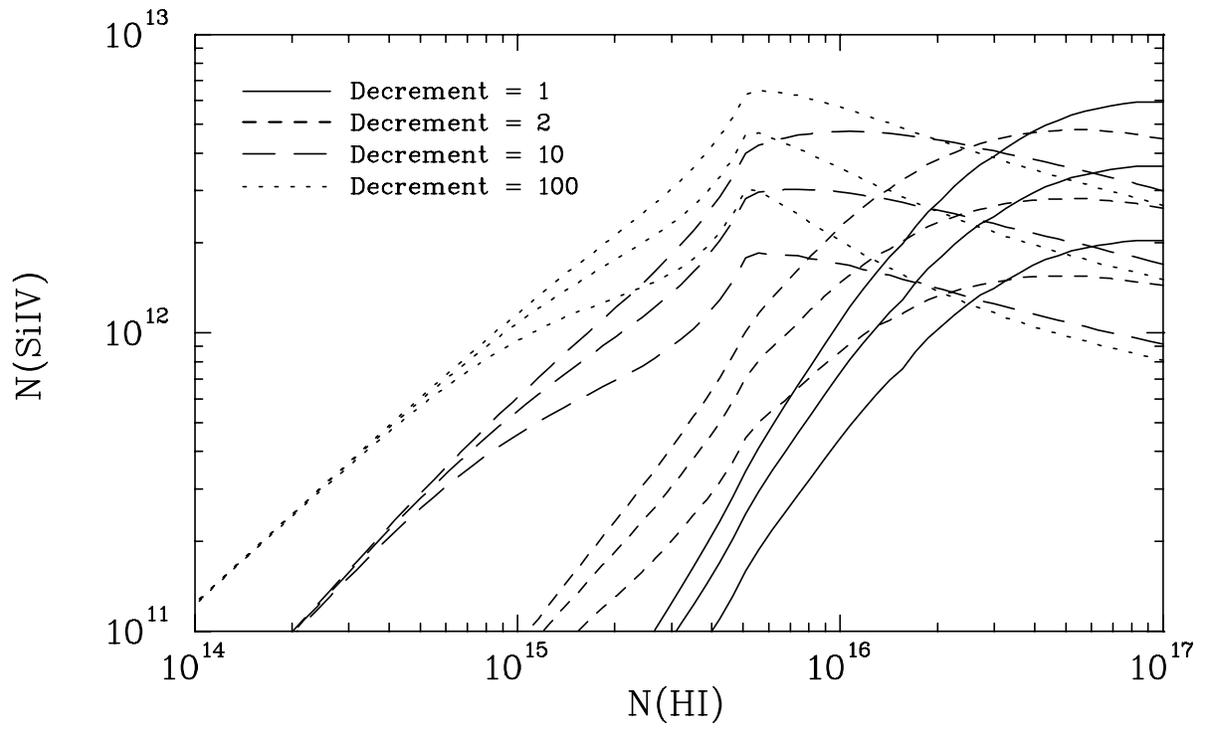}
\caption{Predicted Si~IV column density versus H~I column density.
We have varied the the Si~IV to Si~III DR rate and left the other rates
unchanged.  See Fig.~\ref{fig:NVvHI} for further
details.}
\label{fig:SiIVvHI}
\end{figure}

To simulate the effects of the uncertainties in the DR rates, we have
run CLOUDY with the rates onto N~V, O~VI, and Si~IV decreased by a
factor of 2, unchanged, and increased by a factor of 2.
Figures~\ref{fig:NVvHI} to \ref{fig:SiIVvHI} show the resulting N$_{\rm
NV}$, N$_{\rm OVI}$, and N$_{\rm SiIV}$ versus N$_{\rm HI}$.  The
resulting column densities differ from the column densities predicted
using the unchanged DR rates by factors of up to $\sim 1.9$ for N~V and
Si~IV and $\sim 1.6$ for O~VI.  This translates into a factor of up to
$\sim 1.6-1.9$ uncertainty in any derived abundances.

\begin{figure}
\plotone{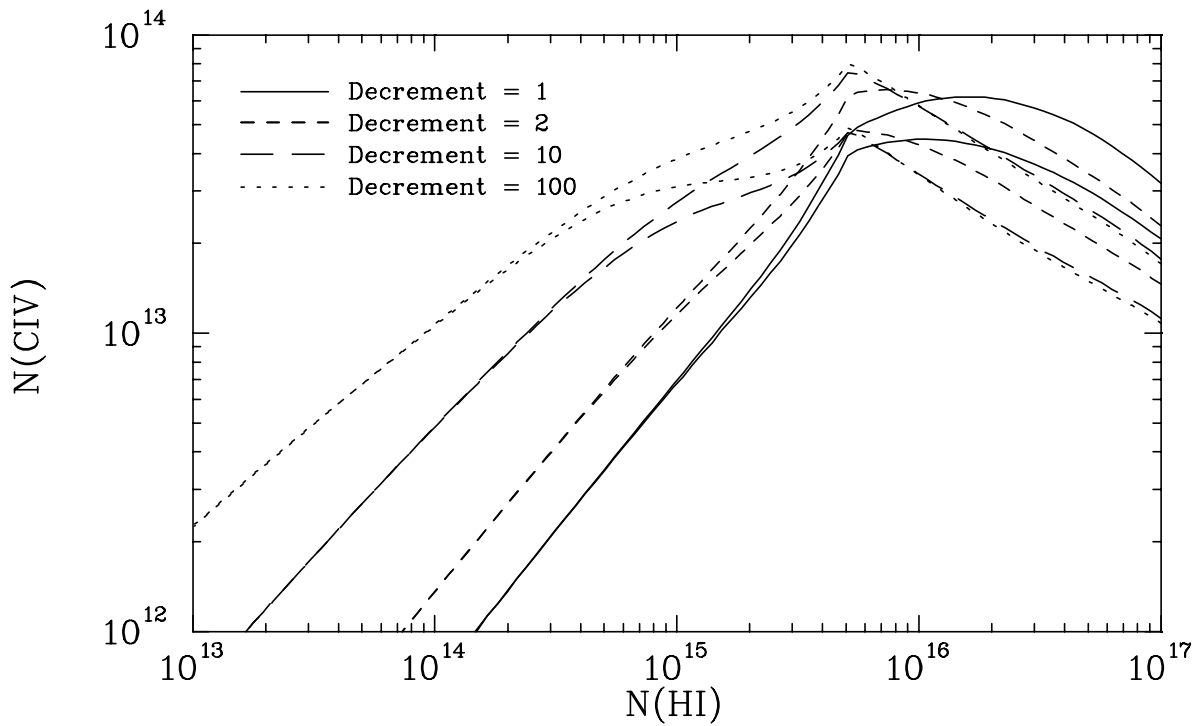}
\caption{Predicted C~IV column density versus H~I column density.  We
have varied the the C~IV to C~III DR rate and left the other rates
unchanged.  For each set of curves, the results are shown for the rate
unchanged (upper curve) and increased by a factor of 2 (lower curve).
See Fig.~\ref{fig:NVvHI} for further details.}
\label{fig:CIVvHI}
\end{figure}

For C~IV, CLOUDY uses the low temperature DR rates of Nussbaumer \&
Storey (1983) and the high temperature DR rates of Shull \& van
Steenberg (1982).  These rates lie at the lower end of the range of
published C~IV DR rates.  We use the C~IV DR rates unchanged and also
increased by a factor of 2.  Figure~\ref{fig:CIVvHI} shows the
resulting N$_{\rm CIV}$ versus N$_{\rm HI}$.  The predicted column
density can be as much as a factor of $\sim 1.7$ smaller than that
predicted using the unchanged DR rates.  This could increase any
inferred abundances by up to a factor of 1.7.

\begin{figure}
\plotone{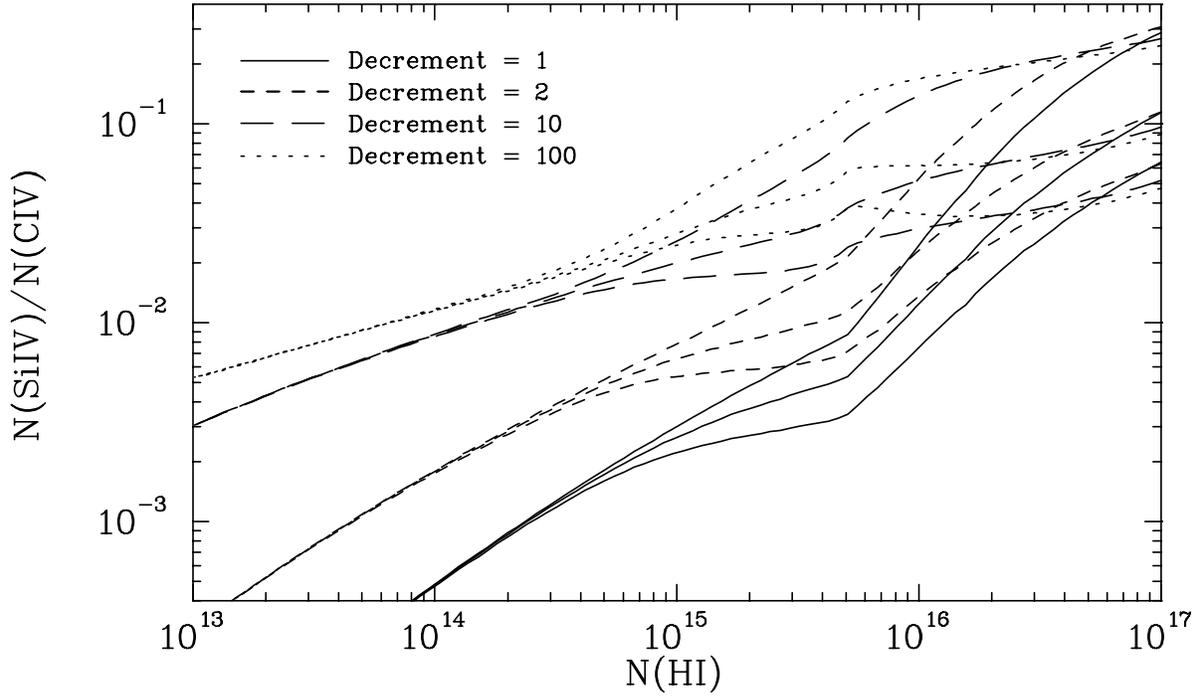}
\caption{Predicted Si~IV to C~IV column densities versus H~I column
density.  We have varied the C~IV and Si~IV DR rates and left the other
rates unchanged.  For each set of three curves, we have decreased the
Si~IV rate by a factor of 2 and increased the C~IV rate by a factor of
2 (upper curve), left both rates unchanged (middle curve), and
increased the Si~IV rate by a factor of 2 while leaving the C~IV rate
unchanged (lower curve).  See Fig.~\ref{fig:NVvHI} for further
details.}
\label{fig:SiIVCIVvHI}
\end{figure}

In Fig.~\ref{fig:SiIVCIVvHI} we have plotted the predicted N$_{\rm
SiIV}$/N$_{\rm CIV}$ ratio versus N$_{\rm HI}$.  Here we vary the SI~IV
and C~IV DR rates.  The resulting ratio could be up to 1.9 times
smaller or 3.1 times larger than the ratio predicted using the
unchanged DR rates.  Hence, the inferred Si/C abundance ratio could be
up to 3.1 times smaller or 1.9 times larger than that inferred using
the unchanged DR rates.

The inferred Si/C ratio for the IGM is used to constrain the IMF of the
earliest generation of stars.  Giroux \& Shull (1997) inferred a
relative abundance ratio for the IGM of ${\rm Si/C} \sim 2({\rm
Si/C})_\odot$.  Results such as those shown in
fig.~\ref{fig:SiIVCIVvHI} indicate that uncertainties in the DR rates
can either make ${\rm Si/C} < ({\rm Si/C})_\odot$ or $>3({\rm
Si/C})_\odot$.  However, Woolsey \& Weaver (1995) have shown that, even
if massive stars dominate the IMF, chemical evolution models with
Si/C$>3({\rm Si/C})_\odot$ are unrealistic.  Abundance ratios this
large would thus suggest that the metagalactic radiation field is not
purely due to AGN but includes a significant component from stellar
radiation (Giroux \& Shull 1997).

\begin{figure}
\plotone{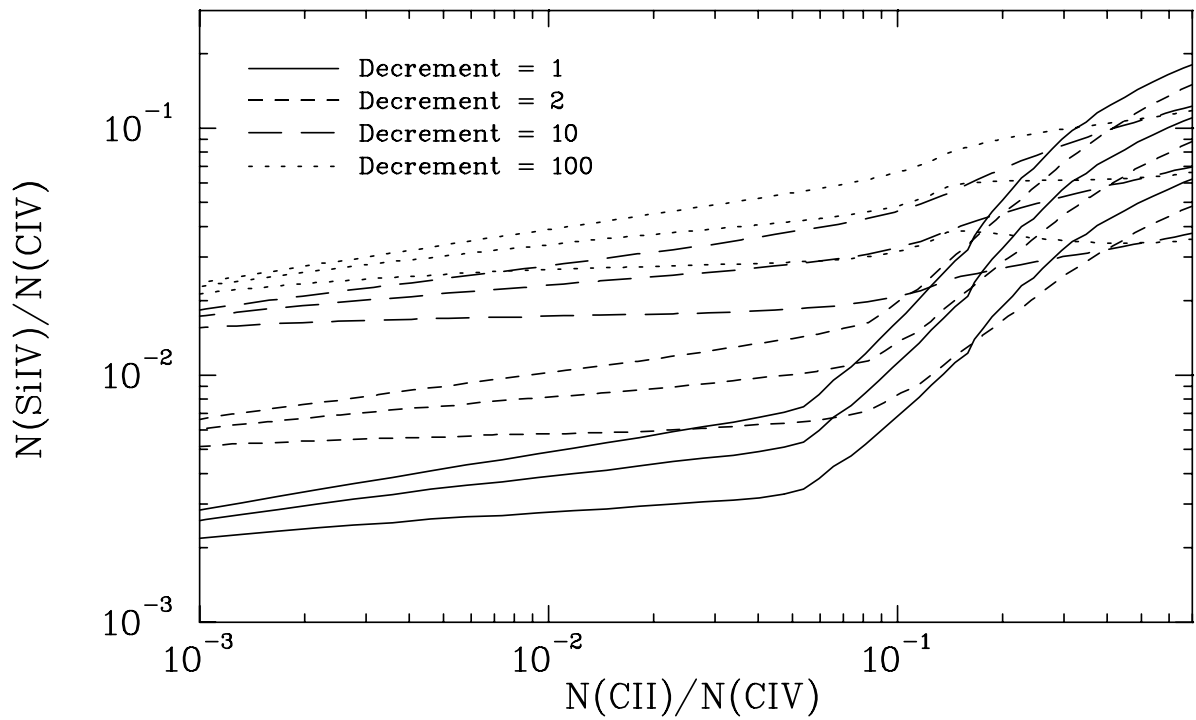}
\caption{Ratio of Si~IV to C~IV column densities versus C~II to C~IV
column densities.  We have varied the the Si~IV to Si~III DR rate and
left the other rates unchanged.  See Fig.~\ref{fig:NVvHI} for further 
details.}
\label{fig:SiIVCIVvCIICIV}
\end{figure}

In Fig.~\ref{fig:SiIVCIVvCIICIV} we have plotted the predicted N$_{\rm
SiIV}$/N$_{\rm CIV}$ versus N$_{\rm CII}$/N$_{\rm CIV}$.  Comparisons
between the observed ratios and model predictions are often used to
constrain the magnitude of the decrement in the radiation field at 4
Ryd.  The magnitude of the decrement affects the amount of He II
photoionization heating of the IGM.  Accurately determining this
decrement has a direct bearing on the issue of late He~II reionization,
which could significantly affect the temperature-density relation of
the IGM, and hence the interpretation of Ly-$\alpha$ forest
observations (Miralda-Escud\'e 1994; Hui \& Gnedin 1997).  Many of the
measured ratios fall in the range of $10^{-2} \lesssim {\rm
N(CII)/N(CIV)} \lesssim 10^0$ (Songaila \& Cowie 1996; Boksenberg 1989;
Songaila 1998).  Our models demonstrate that in this range the
variation in the predicted N$_{\rm SiIV}$/N$_{\rm CIV}$ ratio due to a
factor of 2 uncertainty in the DR rates can be as large as that due
to a factor of 10 change in the decrement.

\section{Conclusions}
\label{sec:Conclusions}

We have shown the effects on IGM models due to the estimated
uncertainties in the DR rates.  These uncertainties limit our ability
to constrain the chemical abundances and the shape of the metagalactic
radiation field at high redshift.  Measurements of the relevant DR
resonance strengths and energies are the only unambiguous way to remove
these atomic physics uncertainties.

\acknowledgements

The author would like to thank G.\ J.\ Ferland and K.\ T.\ Korista for help
with CLOUDY and A.\ Crotts, L.\ Hui, S.\ M.\ Kahn, and F. Paerels for
stimulating discussions.  This work was supported in part by NASA High
Energy Astrophysics X-Ray Astronomy Research and Analysis grant
NAG5-5123.

\vfill
\clearpage
\eject

\clearpage
\eject

\end{document}